\documentclass[conference]{IEEEtran}
\IEEEoverridecommandlockouts

\usepackage[cmex10]{amsmath}

\usepackage{tikz}
\usepackage{adjustbox}
\usepackage{algorithm, algorithmic}             
\usepackage{booktabs}
\usepackage{epstopdf}
\usepackage{enumerate}
\usepackage{pdfpages}
\usepackage{afterpage}
\usepackage{color, colortbl}
\usepackage{footmisc}
\usepackage{etoolbox}
\usepackage{tablefootnote}
\usepackage{makecell}
\usepackage{float}
\usepackage{eucal}
\usepackage{ragged2e} 

\usepackage{xcolor}

\usepackage{mathtools}

\usepackage{mathrsfs}  
\usepackage{pifont}
\usetikzlibrary{patterns}

\ifCLASSOPTIONcompsoc
\usepackage[nocompress]{cite}
\else
\usepackage{cite}
\fi

\usepackage{enumitem}
\setlist{nosep}

\usepackage{amsthm}
\theoremstyle{definition}

\usepackage{graphicx}
\graphicspath{{./figures/}}

\usepackage{array}
\makeatletter
\newcommand{\printfnsymbol}[1]{%
	\textsuperscript{\@fnsymbol{#1}}%
}
\makeatother

\ifCLASSOPTIONcompsoc
\usepackage[caption=false,font=footnotesize,labelfont=sf,textfont=sf]{subfig}
\else
\usepackage[caption=false,font=footnotesize]{subfig}
\fi

\usepackage{multirow}
\usepackage{url}

\newcommand{\minisection}[1]{\vspace{5pt}\noindent{#1.}}

\usepackage{url}

\begin{document}
	
\title{Optimization of Topology-Aware Job Allocation on a High-Performance Computing Cluster by Neural Simulated Annealing}

\author{\IEEEauthorblockN {Zekang Lan$^{1}$, Yan Xu$^{2}$, Yingkun Huang$^{1}$, Dian Huang$^{1}$, Shengzhong Feng$^{1}$}%
	\thanks{$^{1}$National Supercomputing Center in Shenzhen(Shenzhen Cloud Computing Center), CN
		{\tt\small \{lanzk, huangyk, huangdian,fengsz\}@nsccsz.cn}}%
	\thanks{$^{2}$Beijing University of Technology, CN
		{\tt\small yxu@bjut.edu.cn}}%
}

\maketitle

\begin{abstract}
	Jobs on high-performance computing (HPC) clusters can suffer significant performance degradation due to inter-job network interference. Topology-aware job allocation problem (TJAP) is such a problem that decides how to dedicate nodes to specific applications to mitigate inter-job network interference. In this paper, we study the window-based TJAP on a fat-tree network aiming at minimizing the cost of communication hop, a defined inter-job interference metric. The window-based approach for scheduling repeats periodically taking the jobs in the queue and solving an assignment problem that maps jobs to the available nodes. Two special allocation strategies are considered, i.e., static continuity assignment strategy (SCAS) and dynamic continuity assignment strategy (DCAS). For the SCAS, a 0-1 integer programming is developed. For the DCAS, an approach called neural simulated algorithm (NSA), which is an extension to simulated algorithm (SA) that learns a repair operator and employs them in a guided heuristic search, is proposed. The efficacy of NSA is demonstrated with a computational study against SA and SCIP. The results of numerical experiments indicate that both the model and algorithm proposed in this paper are effective.
\end{abstract}


\IEEEpeerreviewmaketitle

\section{INTRODUCTION}
High-performance computing (HPC) cluster is a parallel processing system that employs thousands of interconnected  stand-alone computers to work cooperatively as a single, integrated computing resource. Interconnect topology—which includes the layout of computer nodes (hereinafter referred to as nodes), switches, and channels—plays a key role in determining network performance, which is a critical aspect in HPC clusters. In recent years, inter-job network interference has been identified as a culprit behind job performance variability in HPC systems. For the clusters with different interconnect topologies, e.g., torus-, dragonfly-, and fat-tree based, inter-job network interference can cause production application to slow down by 66\%-150\% \cite{2017Predicting} \cite{2018MitigatingS} \cite{2020Jigsaw} \cite{2020The}.

Topology-aware job allocation problem (TJAP) decides how to dedicate nodes to specific applications to mitigate inter-job network interference. Since the running times of the parallel applications are affected by the frequent communications between the threads of these parallel applications, which exchange data and messages through communication infrastructures such as the message passing interface, the performance of a communication-intensive application is highly dependent on the location of the individual computing nodes that are communicating with each other. Today’s resource and job management system can be configured to support TJAP, for example, SLURM, a well-known and widespread, supports three-dimensional torus interconnect, fat-tree, and dragonfly networks, using slightly different algorithms. However, the existing scheduling system makes decisions in a per-job manner, e.g., First-Come-First-Serve (FCFS), in which each job is dispatched to system resources without considering subsequent jobs. While making isolating job decisions may provide a good short-term optimization, it is likely to result in poor performance in the long term. Differing from the per-job mode that allocates jobs to nodes immediately upon arrival, the window-based approach collects the jobs in a queue until a specified condition is met \cite{2012Integer}\cite{2014Balancing}. The window-based assignment step is repeated periodically. At each period, one assignment problem that maps jobs to available resources has to be solved. Usually, integer programming techniques are used. Compared with a per-job manner, the window-based approach takes the jobs in the waiting queue into comprehensive consideration which leads to a better job allocation solution. Nevertheless, as a window-based system would need to explore a larger solution space, thus a more efficient algorithm is needed.

Typically, TJAP problem is quite similar to quadratic assignment problem (QAP), which has been proven to be NP-Complete. Consequently, some heuristic algorithms have been proposed, such as genetic algorithm \cite{2002A} and evolution algorithm \cite{1992A}. In order to respect the tight time frames imposed by real-time decision-making, one commonly used remedy is a contiguous allocation strategy in that the scheduler assigns each job a compact and contiguous set of computing nodes. This strategy benefits application’s performance by maintaining the locality of allocated nodes and decreasing network contention brought on by multiple jobs running simultaneously and sharing network bandwidth. Based on the contiguity rule, we don’t need to take the allocations of all nodes as decision variables. Instead, only the allocation of one of requesting nodes has to be taken as a decision variable, while the allocations of other nodes will be determined following the contiguous rule. As a result, the formulation of TJAP is not a complex QAP, but linear programming.

In this paper, we study the window-based TJAP on a fat-tree network. The inter-job network interference is measured by a defined communication-hop (CH) cost. A special continuity allocation strategy is proposed, which is to continuously allocate the requested nodes of jobs based on a one-dimensional sequence of idles nodes. According to whether the sequence will be updated when one job is allocated, the continuity allocation strategy is further divided into two situations: static continuity assignment strategy (SCAS) and dynamic continuity assignment strategy (DCAS). The paper’s contribution is twofold. First, we formulate the TJAP based on the SCAS as an integer programming problem. Secondly, we develop a new algorithm, i.e., neural simulated annealing (NSA), for the TJAP based on the DCSA.

The rest of this paper is organized as follows. In Section \uppercase\expandafter{\romannumeral2}, we describe the work related to job scheduling of HPC, especially in TJAP. In Section \uppercase\expandafter{\romannumeral3}, the TJAP is introduced in detail. In Section \uppercase\expandafter{\romannumeral4}, a model based on the SCAS is given. In Section \uppercase\expandafter{\romannumeral5}, a new algorithm named NSA is presented to solve the TJAP based on the DCAS. In Section \uppercase\expandafter{\romannumeral6}, the results obtained are given. Finally, we conclude the paper in Section \uppercase\expandafter{\romannumeral7}.

Source code associated with this paper is available at: https://github.com/lzk23/hpcschedule.

\section{RELATED WORK}

Strictly speaking, job scheduling in a scheduler system contains two parts, namely job prioritizing and job allocation. Job prioritizing makes a decision about the order in which jobs are allowed to run. FCFS and EASY-backfilling (BF) are very simple examples of priority scheduling algorithms. BF is an approach that may backfill (execute a job out of order) whenever there are enough idle nodes, and the backfilled job would not delay the job at the head of the queue. About 90\%$\sim$95\% of job schedulers use BF \cite{2007Backfilling}. Job scheduling mainly considers several factors including fairness \cite{2007Backfilling}, energy \cite{2013Reducing}, utilization \cite{2020Jigsaw}, and performance. This paper focuses on TJAP, which is used to alleviate inter-job interference and improve performance.

Most of the literature on TJAP adopts the per-job manner. Kaplan et al. \cite{2013Optimizing} proposed a job allocation methodology to jointly minimize the communication cost of an HPC application while also reducing the cooling energy. Georgiou et al. \cite{inproceedings}\cite{2018Large} considered the applications communication matrix. Yan et al. \cite{2019LPMS} proposed a label-propagation-based process mapping method, namely LPMS, that is both low-cost and fits well in shared HPC systems (one node can be shared by multiple jobs). Ryu et al. \cite{2020Towards} present a topology-aware deep reinforcement learning scheduler that simultaneously selects jobs and assigns resources to them. The machine bandwidths and current GPU to job assignments representing the cluster topology are presented as input to the deep neural network. Smith et al. \cite{2020Jigsaw} developed a new job-isolating scheduling approach for three-level fat-trees that eliminates inter-job network interference while maintaining high system utilization. To analyze the performance of network topologies and their different configurations prior to building them, network designers turn to analytical modeling and simulation, such as \cite{2016Enabling} and \cite{2018Joint}.

Few works on TJAP are window-based. Soner et al. \cite{2012Integer} proposed a scheduling algorithm AUCSCHED that attempts to allocate contiguous blocks on one-dimensional array of nodes. CPLEX is used to solve the 0-1 integer model. Yang et al. \cite{2014Balancing} maintained a list of slots to preserve node contiguity information for resource allocation. A 0-1 multiple knapsack model was formulated and solved by branch-and-bound and greedy algorithms, respectively.

\section{PROBLEM STATEMENT}
In this section, we first briefly introduce the fat-tree network topology and detail the inter-job interference metric (i.e., communication-hop cost). Then, we describe two different continuous allocation strategies, namely SCAS and DCAS. Finally, we formally introduce the TJAP.

\subsection{Fat-tree network and inter-job interference metric}
The fat-tree network is a universal network for provably efficient communication. The common type of fat-tree used in most high-performance networks is $k$-ary fat-tree. In particular, for a given radix of $k$ (the number of ports in one switch), there are $k$ pods in the fat-tree topology. In each pod, there are $k$ switches evenly distributed in two layers and each switch has $k$ ports. Each switch at lever $L_0$ is connected to $k/2$ nodes and $k/2$ switches at level $L_1$ (1 port for each). Similarly, each switch at level $L_1$ is connected to $k/2$ switches at level $L_0$ and $k/2$ switches at level $L_2$. Therefore, the fat-tree network has a full bisection bandwidth. Consequently, each pod connects with $(k/2)\cdot(k/2)$ nodes and $k\cdot\left(k/2\right)\cdot\left(k/2\right)=k^3/4$ nodes are interconnected in the system. Fig. 1(a) shows a 4-ary fat-tree, where there are 4 nodes in each pod and 16 nodes in total.

For a real system with a certain number of nodes, the full bisection fat-tree topology needs to be revised. Hence, pruned fat-tree configurations are commonly embraced in production systems. Based on the $k$-ary fat-tree network model, some pods are pruned to match the real system. For example, a pruned fat-tree model with two pods corresponding to the full bisection fat-tree in Fig.1(a) is illustrated in Fig. 1(b).

\begin{figure*}
	\begin{center}
		\begin{tabular}{cc}
			\includegraphics[width=.45\textwidth,angle=0]{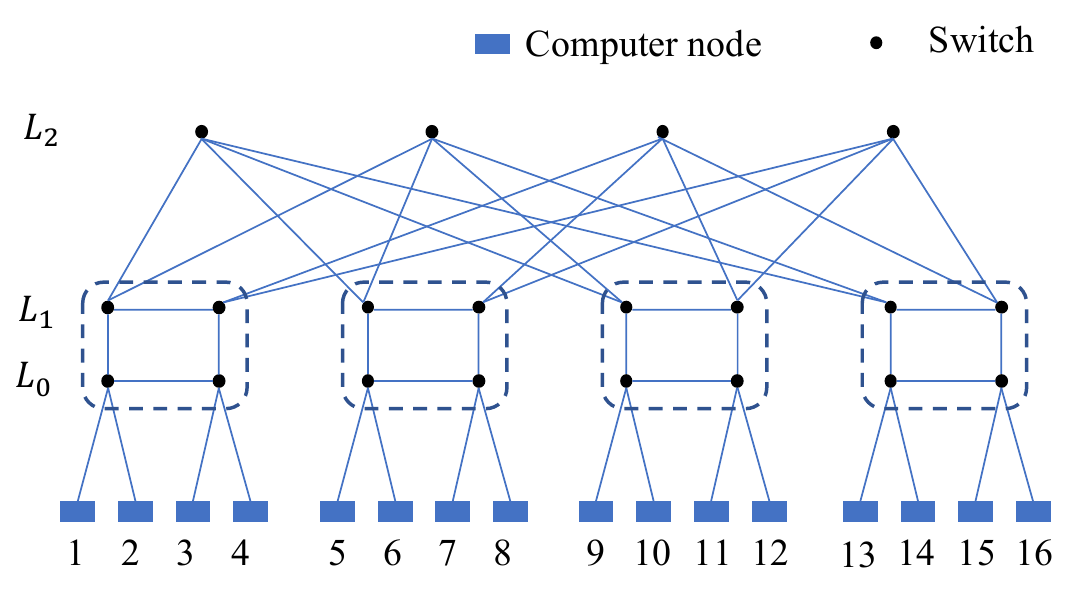}
			&\includegraphics[width=.45\textwidth,angle=0]{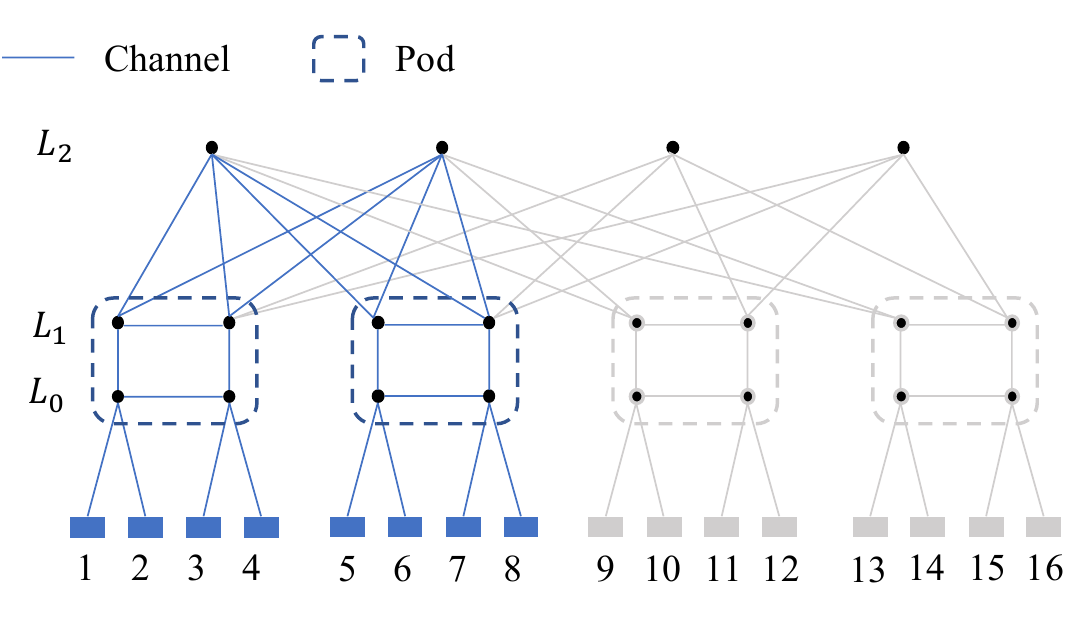}\\[0pt]
			(a)  k-ary fat-tree architecture     & (b) Pruned fat-tree configuration\\
		\end{tabular}
	\end{center}
	\vskip-10pt
	\caption{Fat-tree network topology.}
	\label{fig:1}
\end{figure*}

As mentioned before, the fat-tree network topology may suffer from serious inter-job interference that hurts performance. In a shared HPC system, the inter-job interference occurs in a node. In this paper, we assume that all jobs request more than one node. The jobs requesting exactly one node are not considered since they have no inter-job interference. Consequently, nodes are exclusive, that is, once one node has been occupied by one unfinished job, it cannot be allocate to any other jobs. In a result, the inter-job interference does not occur inside a node. 

Intuitively, in order to reduce inter-job interference, jobs should be placed in as few pods as possible. This paper introduces a customized CH cost to measure the inter-job interference. A CH is an intermediate connection linking two devices (i.e., nodes or switches). For two appointed nodes in a fat-tree network, if they have the same parent’s switch (connected directly with these two nodes in level $L_0$), the number of CHs is 2; otherwise, if they are in the same pod, the number of CHs is 4; otherwise, if they are in different pods, the number of CHs is 6. Given a set of nodes $N_j\left(N_j\geq2\right)$ allocated to job $j$, the formular for computing the cost of CH is given as:
\begin{equation}
	C(N_j)=\frac{1}{n_j}\sum_{i \in N}\sum_{k \in N:i \ne k}ch_{i,k} 
\end{equation}
where $n_j$ is the number of requested nodes by job $j$; $c$ is the cost for one CH; $h_{i,k}$ is the number of CHs between nodes $i$ and $k$.

\subsection{Continuity allocation strategy}\label{3.2 continuity allocation strategy}
To avoid the TJAP falls into a QAP which is hard to be solved, a special continuity allocation strategy is employed. To illustrate, nodes in the fat-tree network are labeled with unique identifications based on a linear order from left to right. At each scheduling period, all idle nodes are arranged in non-descending order of their identifications to form a sequence $Q$. Based on sequence $Q$, we have to decide the first allocation node for each job. Once the first allocation node is decided, the remaining nodes are allocated consecutively following the order of nodes given in sequence $Q$. In particular, if it has not enough $n$ continuous nodes starting from node $i$, the remaining nodes are taken from the head of sequence $Q$. Following this rule, the nodes in the same pod are likely to be assigned to one job thus potentially reducing CH costs. 

Depending on whether it is allowed to update sequence $Q$ when one job is allocated, the continuity allocation strategy is classified into two categories: SCAS and DCAS. In the SCAS, sequence $Q$ will not be updated when one job is allocated. While in the DCAS, sequence $Q$ will be updated dynamically by removing the assigned nodes once a job is allocated. Each job is allocated continuously based on a new sequence of idle nodes. For example, for the 4-ary fat-tree cluster in Fig. 1, considering that only nodes 1-10 are idle, that is, $Q$= {1-10}. Two jobs called job 1 and job 2 need to be allocated, and both of them request 4 nodes. Job 1 is assigned to nodes 5-8. For the SCAS, there are three feasible solutions as illustrated in Fig. 2(a) for job 2. For the DCAS, sequence $Q$ will be updated as $Q$= \{1-4,9,10\} after job 1 is allocated to nodes 5-8, there are six feasible solutions with three solutions in Fig.2(a) and three solutions in Fig.2(b) for job 2. That is, the solutions for the SCAS are also feasible for the DCAS. 

\begin{figure}
	\begin{center}
		\begin{tabular}{cc}
			\includegraphics[width=.45\textwidth,angle=0]{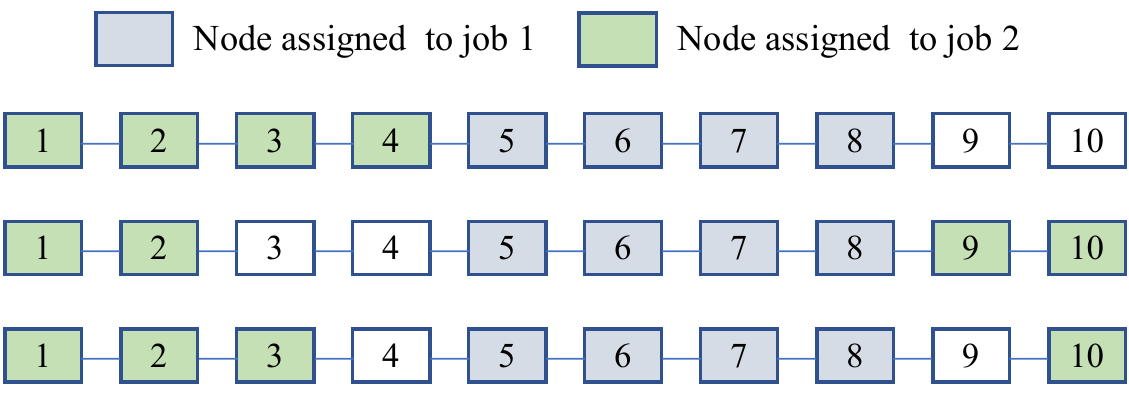}\\[0pt]
			(a) Static continuity assignment\\
			\includegraphics[width=.45\textwidth,angle=0]{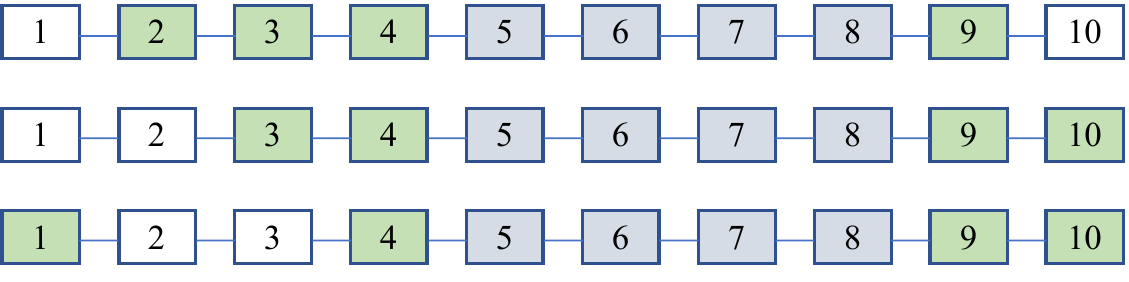}\\[0pt]
			(b) Dynamic continuity assignment\\
		\end{tabular}
	\end{center}
	\vskip-10pt
	\caption{Static and dynamic continuity assignment strategies.}
	\label{fig:2}
\end{figure}

\subsection{Description of the TJAP}\label{3.3. Description of the TJAP}
Our TJAP is based on a fat-tree network. The objective of the TJAP is to assign a set of nodes to each job according to the SCAS or the DCSA to optimize the CH cost as shown in Eq. (1). More specifically, we assume a dynamic job arrival, i.e., a variable number of jobs waiting at the input queue and more jobs keep coming with the progress of time. Our TJAP is window-based that indicating the TJAP involves periodically taking the jobs in the queue and solving an assignment problem that maps jobs to the available nodes. The length of each time period is given by $\tau$(e.g., 30s). That is, the scheduling approach must take a decision in an online fashion at every time period. Each job $j$ has an arrival time, a requested number of nodes, and a processing time (the user submits an estimated processing time; the true processing time is known when the job completes). Job $j$ is only known to the scheduler after the arrival time when it is dispatched from the scheduler. Preemption and migration are not allowed, that is, once a job is begun it must be executed to completion on the same set of nodes. 

Note that, it’s possible that the available idle nodes are insufficient to fulfill the requirement of all jobs in the waiting queue in one time period. Thus, besides job allocation, we also have to decide on job prioritizing. In particular, we assign each job an attribute, named “number of waiting periods” (NWP), whose value is initialized to 0. For jobs that are not assigned in the current period, the NWP values will be increased by 1. At each decision period, jobs with larger NWP values have higher priorities. For jobs with the same NWP values, jobs with fewer number of requested nodes have higher priority. According to the defined priority rules, as many jobs as possible are selected to allocate. The remaining unassigned jobs will wait for allocation in the next time period. 

\section{STATIC CONTIGUITY ASSIGNMENT}\label{4. static contiguity assignment}
In this section, we first formulate the TJAP based on the SCAS in Section \uppercase\expandafter{\romannumeral4}-A. Then, the NP-hardness of the model is proved in Section \uppercase\expandafter{\romannumeral4}-B. 
\subsection{Formulation of the TJAP based on the SCAS} \label{4.1 formulation of the TJAP based on the SCAS}
To formulate the model based on the SCAS, we first make the definitions as follows: $J$, set of jobs that are picked from the waiting queue as described at the end of Section \uppercase\expandafter{\romannumeral3}-C; $N$, set of idles nodes. Accordingly, a node sequence $Q$ can be constructed based on nodes in $N$ with their identifications increasing; $n_j$, the requested number of nodes by job $j$; $D\left(i,n\right)$, the $n$ continuous nodes starting from node $i$ in sequence $Q$; $l_{j,i}$, the CH cost for the nodes in $D\left(i,n_j\right)$; $y_{j,i}$, binary variable indicating whether job $j$ selects node $i$ as its first allocation node. The following gives the model based on the SCAS, denoted as $\mathrm{M}_{\mathrm{SCAS}}$:
\begin{equation}
     (\mathrm{M}_{\mathrm{SCAS}}) \mathrm{min} \sum_{j \in J}\sum_{i \in N}l_{j,i}y_{j,i}
\end{equation}
\begin{equation}
	\sum_{i \in N}y_{j,i}=1, \, \forall j \in J
\end{equation}
\begin{equation}
	y_{j,i} \le x_{j,k}, \, \forall j \in J,i \in N, k \in D(i,n_j)
\end{equation}
\begin{equation}
	x_{j,i},y_{j} \in \{0,1\}, \, \forall j \in J,i \in N
\end{equation}
where the objective function (2) represents minimizing the total CH cost. Constraint (3) restricts that each job selects exactly one first allocation node. Constraint (4) ensure that if node $i$ is selected as the first allocation node of job $j$, then all the nodes in $D\left(i,n_j\right)$ are assigned to job $j$. Constraint (5) set the domain of variables.

\subsection{NP-hardness of the model based on the SCAS}\label{NP-hardness of the model based on the SCAS}
The above model can actually be converted into a constrained shortest path problem (CSP), which has been proved to be NP-Completed \cite{2001Constrained}. The equivalence of the TJAP and CSP is proved later.

Given an idle node sequence $Q=\{1,2,\ldots,i,\ldots,\ n\}$ and $J=\{0,1,..,j,\ldots,\left|J\right|\}$. The set of number of requested nodes for jobs is $R$. The number of jobs that request $r$ nodes is $\varepsilon_r$, thus $\left|J\right|=\cup_{r\in R}\varepsilon_r$. Construct a directed graph $G=\left(V,E\right)$, where $V$ is the node set corresponding to the nodes in $Q$, and $E$ is the arc set. The arc set $E$ encompasses two types of arcs: (1) the first is the transition arcs, which refer to the directed arcs connecting any node to the next first node according to sequence $Q$; (2) the second is the operation arcs, which are the connections from node $i$ to its next $(r-1)$-th node (if any) for each node $i\ \in V$ and each $r\ \in R$. For an operation arc that connects node $i$ to its next $(r-1)$-th node, the set of its span nodes is $\{i,i+1,\ldots,\ i+r-1\}$. Note that, the operation arc returning “backwards” to the head of nodes in sequence $Q$ is not considered for brevity. The set of operation arcs spanning $r$ nodes is $E_r$. The weight of any adjacent arc is associated with 0, and the weight of an operation arc is associated with the CH cost of the nodes such that spanned by the arc. Then, the TJAP can be transformed into a CSP that finds the shortest path from node 1 to node $n$ in $G$, meeting the requirement of passing through $\varepsilon_r$ operation arcs in $E_r$. After the shortest path is determined, the allocation nodes for job $j$ are those nodes spanned by any operation arc that are both in the shortest path and in $E_{n_j}$. 

For example, suppose that $Q=\{1-10\}, J=\{0,1\}$. The number of requested nodes by any job is 4, that is, $R=\{4\}$ and $n_4=2$. The corresponding graph $G$ is given in Fig. 3. Since all jobs requesting for 4 nodes, the operation arcs are constructed by connecting each node to its next third node (if any). For example, there is an operation arc that connects from node 1 to node 4, and the set of its span nodes is \{1,2,3,4\}. The TJAP is then converted into the CSP that finds a shortest path from node 1 to node 10 such that passes through the operation arcs exactly twice. 
\begin{figure}[]
	\centerline{
		\includegraphics[width=0.95\columnwidth]{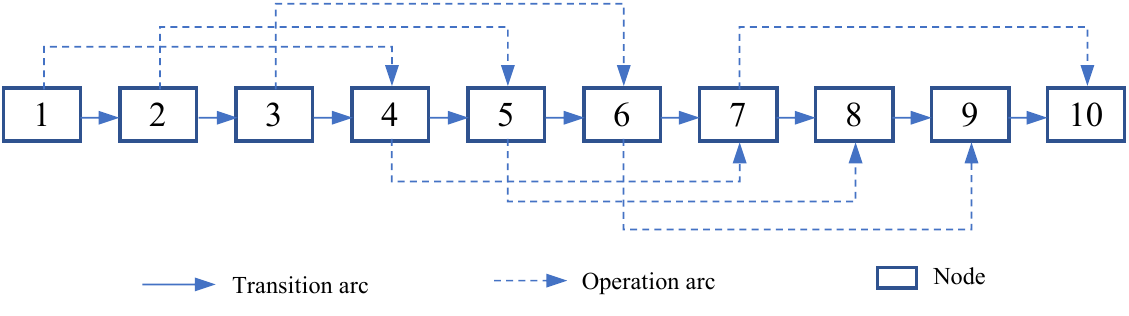}}
	\caption{An illustration for converting the TJAP to a CSP.}
	\label{fig:repair_problem}
\end{figure}

\section{DYNAMIC CONTIGUITY ALLOCATION}
Considering that the model based on the DCAS is represented as $\mathrm{M}_{\mathrm{DCAS}}$. Intuitively, the feasible solution space of $\mathrm{M}_{\mathrm{SCAS}}$ is included in that of $\mathrm{M}_{\mathrm{DCAS}}$. Therefore, $\mathrm{M}_{\mathrm{DCAS}}$ is also NP-hard. In addition, $\mathrm{M}_{\mathrm{SCAS}}$ can be directly solved by a solver (e.g., CPLEX, GUROBI), whereas $\mathrm{M}_{\mathrm{DCAS}}$ is a multi-stage optimization problem, which cannot be solved directly by a solver.

In order to solve $\mathrm{M}_{\mathrm{DCAS}}$, we propose a new algorithm called neural simulated annealing (NSA), which is an extension to the simulated annealing (SA) \cite{1983Optimization} that automates the complex task of generating neighbor solutions using reinforcement learning (RL). SA algorithm has been extensively developed and widely used in many optimization problems. It can avoid getting trapped in a local optimum and attain better solutions by accepting worse solutions with a certain probability. Our method is inspired by \cite{2019Neural}, who presented a neural large neighborhood search algorithm that learns repair operators for vehicle routing problem.

\subsection{Overall algorithm of NSA}
The overall NSA algorithm is described in Algorithm 1. Aside from the RL algorithm used at Step 4(b), NSA has a same progress as SA. Thus, the progress of NSA is also controlled by three parameters $T^{\mathrm{max}}$, $T^{\mathrm{min}}$, and $t^{\mathrm{max}}$: $T^{\mathrm{max}}$ is the maximal temperature; $T^{\mathrm{min}}$ is the minimal temperature; $t^{\mathrm{max}}$ is the maximal iteration steps. The loop continues as long as the iteration times reach $t^{\mathrm{max}}$. At iteration $t$, the temperature $T$ decreases to $T^{\mathrm{max}}e^\frac{rt}{t^{\mathrm{max}}}$, where $r$ is a cooling factor and calculated as $-\mathrm{log}T^{\mathrm{max}}/T^{\mathrm{min}}$. With the reduction of $T$, the acceptance rate for a worse solution will also decrease as shown at Step 8 in Algorithm 1. 

\begin{algorithm}
	\footnotesize
	\textbf{Input:} Cost function $\textsc{Cost()}$, Maximal(Minimal) temperature $T^{\mathrm{max}}$ $(T^{\mathrm{min}})$, Maximal iteration steps $t^{\mathrm{max}}$  \\
	\textbf{Output:} The best found feasible solution $x_{\mathrm{best}}$\\
	\begin{algorithmic}[1]
		\STATE Set $r=-log(T^{\mathrm{max}}/T^{\mathrm{min}})$
		\STATE Generate an initial solution $x$ by a heuristic algorithm and memorize $x$ as $x_{\mathrm{best}}$
		\WHILE{$t \le t^{\mathrm{max}}$}
			\STATE Randomly generate $x^\prime$ from $x$: (a) remove a random number of jobs from the assignment solution of $x$; (b)repair the destroy solution by RL algorithm
			\IF{$x^\prime$ is better than $x$}
				\STATE $x \gets x^{\prime}$
			\ELSE
				\STATE $x \gets x^{\prime}$ with probalility of $e^{[\textsc{Cost}(x)-\textsc{Cost}(x^\prime)]/T}$
			\ENDIF
			\STATE Memorize $x$ as $x_{\mathrm{best}}$ if $\textsc{Cost}(x^\prime) \textless \textsc{Cost}(x)$
			\STATE $T \gets T^{\mathrm{max}}e^{\frac{rt}{t^{\mathrm{max}}}}$
		\ENDWHILE
		\STATE \algorithmicreturn{} $x_{\mathrm{best}}$
	\end{algorithmic}
	\caption{NSA for the TJAP\label{alg:single}}
\end{algorithm}

More specifically, let $x$ be an initial solution generated by a heuristic algorithm, e.g., a sequential assignment that schedules jobs one by one using the SCAS. SA generates a neighboring solution $x^\prime$ of $x$ by applying a move operator. The move operator can be seen as a combination of a remove operator followed by a repair operator. The remove operator deletes parts of the solution. Specially, given a maximal remove number of jobs as $n^D (n^D\geq1)$, a random number is generated between $[1, n^D]$ as the number of jobs needs to be removed. The repair operator then fixes the destroyed solution by completing unassigned jobs to available nodes based on the DCAS, thus creating a feasible solution $x^\prime$. In particular, the repair operator first sorts all jobs that have been removed by the destroy operator based on their required number of nodes, and then reinsert those jobs in a sequential fashion based on the DCAS. An acceptance criterion that is Metropolis criterion is used to determine whether the new solution should be accepted, i.e., whether $x^\prime$ should replace $x$. After updating $x$ (or not), the search continues until a termination criterion is reached.

The performance of SA heavily depends on the quality of the destroy and repair operator developed by domain experts. While even simple remove operators that destruct parts of a solution purely at random can lead to good solutions, the repair operator often requires the implementation of a sophisticated heuristic. The only difference between NSA and SA is the repair operator. The repair operator in NSA corresponds to a learned parameterization $\theta$ of our proposed neural network (NN) model that repairs a solution in a sequential repair process. During the training of a repair operator, the corresponding model is repeatedly presented with incomplete solutions that have been destroyed by a particular destroy operator. The objective of the training is to adjust the model parameters $\theta$ so that the model constructs high-quality solutions. This capability is of particular importance for problems where instances with similar characteristics are frequently solved in practice. This also means that NSA can be retrained in the case that the characteristics of the encountered instances change, avoiding significant human resources needed to design new operators. 

\subsection{Learning to repair solutions}
The problem of repairing a destroyed solution can be formulated as a RL problem in which an agent (a learned model) interacts with an environment (an incomplete solution) over multiple discrete time steps. In each time step the model picks one unallocated job from $J$ and output the first allocation node for this job. Then the job is allocated following the continuity allocation strategy based on sequence $Q$. After allocation, sequence $Q$ will be update according to the DCAS. The process is repeated until there are no unallocated jobs in $J$. The scheduling sequence of jobs in $J$ can be random or using other methods. In this paper, we sort jobs in $J$ in descending by their requested number of nodes and then select orderly. In the following, we first provide some general mathematical for RL. Then, the subsections concentrate on the TJAP and detail the definition of each component in RL.

In RL, the environment is usually modeled by a Markov decision processes (MDPs), which are discrete-time stochastic control processes used to model decision making. Formally, an MDP is described as a 4-tuple $M=\left(S,A,P_a,R_a\right)$ where $S$ is a finite set of states; $A$ is a finite set of actions; $P_a\left(s,s^\prime\right)$ is the probability that state $s$ changes to $s^\prime$ when action $a$ is taken (because the environment in our case is deterministic, the probability is set to 1); and $R_a\left(s,s^\prime\right)$ is the reward we receive after switching from state $s$ to $s^\prime$.

The objective of a RL training algorithm is to learn a policy $\pi$ that associates each state with a specific action in order to maximize the expected cumulative reward. For a state $s_0$, we can compute the value  $V^{\pi}$ of a policy $\pi$ given by the formula:

\begin{equation}\nonumber
	V^\pi\left(s_0\right)=E^\pi\left[\sum_{t=0}^{+\infty}{\gamma^tR\left(s_t,a_t\right)}\right]
\end{equation}

\noindent where $V^\pi\left(s_0\right)$ denotes the expectation over the distribution of the admissible trajectories $(s_0, a_0, r_0, s_1, a_1, r_1, …)$ obtained by sampling the actions from the policy $\pi$; $\gamma$ denotes the discount-rate, a hyper-parameter that regulates how far into the future the agent looks. A policy is optimal if no other policy produces a higher return value.

\subsubsection{State Representation}
Recall that the repair operator actually schedules jobs one by one (see the beginning of Section \uppercase\expandafter{\romannumeral5}-B). Considering that the current time step in MDP is $t$, and the idle node sequence is updated as $Q$ following the DCAS. Let $n^Q$ be the number of nodes in $Q$. Roughly, the state representation can be a $\left(n^Q\times\beta\right)$ matrix, where $\beta$ is the number of attributes for each node in $Q$. However, sequence $Q$ will be dynamically updated according to the DCAS. When the model architecture of RL contains fully-connected layers (see later), the state size should be fixed. To deal with this, our state representation is a $\left(q\times\beta\right)$ matrix, where $q$ is a given constant parameter. If $n^Q\geq q$, the first $q$ nodes in $Q$ are taken into consideration for state representation; if $n^Q<q$, all the nodes in $Q$ are selected, and the rest elements are filled with 0.

Each node in $Q$ has 4 attributes (i.e., $\beta$=4): (1) The parent’s switch. (2) The pod this node belongs to. (3) The info for the requested number of nodes for the current scheduling job. The current scheduling job is a such that RL needs to decide its first allocation node at the current time step $t$. (4) The info for the total requested number of nodes for the other unallocated jobs. In particular, for the first two attributes, the corresponding values in the state matrix can be given as the normalized identifications for the parent’s switch and pod, respectively. For the last two attributes, if the node is idle, the corresponding values in the state matrix can be formulated as $n_j/n^{\mathrm{max}}$ and $\sum_{j^\prime \in J^\prime\backslash\{j\}}n_{j^\prime}/\left[\left(\left|J^\prime\right|-1\right)n^{\mathrm{max}}\right]$, respectively, where $n^{\mathrm{max}}$ denotes the maximal requested number of nodes for jobs; $j$ refers to the current scheduling job; $J^\prime$ represents the set of unallocated jobs in $J$; otherwise, the corresponding values are set as 0. 

\subsubsection{Action Selection}
For each scheduling job, the action is to select its first allocation node. Once the first allocation node is determined, the allocation nodes for this job are prescribed following the continuity allocation strategy based on sequence $Q$. The action space of RL is defined by $A=\{0,1,\ldots,q-1\}$. When $n^Q<q$, the actions in $\{n^Q,\ldots,q-1\}$ are illegal. Different techniques can be applied to skip illegal actions. One method is to provide a negative reward if the agent takes one illegal action and hopes that it will learn to identify them. This method produces poor results because it complicates the problem by requiring the agent to learn to distinguish between a legal/illegal action and a good/bad action at the same time.

Our method involves using a mask on the output of a NN to transform illegal action values into a small negative number, i.e., the smallest representable number. When we use the softmax function, the probability of illegal action approaches zero. This technique has been studied previously and appears to produce the best results\cite{2020ACloser}. 

\subsubsection{Reward Function}
Recall that the number of jobs removed by the remove operator is generated randomly between $[1, n^D]$. If the random number is $n$, then $n$ jobs need to be scheduled by the repair operator. For the first $(n-1)$ jobs, the reward is given 0, for the $n$-th job, the reward is calculated as the inverse of the CH cost for the $n$ jobs. The opposite value is taken because the larger CH cost means a smaller reward.

\subsubsection{Model Architecture}

\minisection{Fully-Connected Network}
A fully-connected neural network (FCN) \cite{2020Impact} is made up of several fully-connected layers that connect every neuron in one layer to every neuron in the other. The main advantage of FCN is that they are “structure agnostic”, which means that no special assumptions about the input are required. While the fact that FCN is structure agnostic makes it very broadly applicable, such networks tend to perform worse than special-purpose networks tuned to specific problems.

\minisection{Convolutional Neural Network}
The FCN does not use the location information between pixels. For image recognition problems (HPC clusters can be regarded as images and nodes can be regarded as pixels of images), each pixel is closely related to its surrounding pixels, and is less related to pixels far away. If a neuron is connected to all the neurons in the upper layer, after learning the weights of each connection, it may eventually be found that a large number of weights have very small values, that is, these connections are actually insignificant, which undoubtedly makes training very inefficient. Convolutional Neural Network (CNN) architecture \cite{2001Constrained} explicitly assumes that the input is an image, which allows pixel position attributes to be encoded into the model architecture. CNN adopts local connections, and each neuron is no longer connected to all neurons in the upper layer, but only to a small number of neurons. At the same time, CNN adopts weight sharing. A group of connections can share the same weight, rather than each connection having a different weight, which reduces the parameter size.

\subsubsection{Proximal policy optimization}
We use proximal policy optimization (PPO) \cite{2017Proximal} as the RL algorithm to train our agents. PPO implements an actor-critic architecture, which is at the intersection of the policy-based and the value-based RL such that the policy (actor) is optimized in the direction suggested by the value function (critic). The fact that update steps computed at any specific policy $\pi_\theta$ only guarantee predictiveness in a neighborhood around $\theta$ is a fundamental property of policy-based algorithms \cite{2002Deterministic}. In the parameter space, policy-based algorithms keep policy updates close. However, even seemingly minor differences in the parameter space can have a significant impact and degrade policy performance. PPO strives to make the greatest possible improvement while not going so far as to cause performance collapse. It accomplishes this through a series of stochastic gradient ascent steps on a clipped objective function. In this case, the loss function $L$  is given by:

\begin{align}\nonumber
	& L = \min\left(
	\frac{\pi_{\theta}(a|s)}{\pi_{\theta_k}(a|s)}  D^{\pi_{\theta_k}}(s,a), \;\;
	g(\epsilon, D^{\pi_{\theta_k}}(s,a))
	\right)
\\\nonumber
	& g(\epsilon, D) = \left\{
	\begin{array}{ll}
		(1 + \epsilon) D & D \geq 0 \\
		(1 - \epsilon) D & D < 0
	\end{array}
	\right.
\end{align}

\noindent where ${\theta}$ and ${\theta}_k$ are the parameters of the new and the old policies, respectively; ${\varepsilon}$ is a (small) hyper-parameter that roughly states how far away the new policy can deviate from the old one; $D^{{\pi}_{\theta}}$  is an estimate of the advantage function.

\section{Result}
In this section, we describe the experiments used to validate the proposed model and assess the performance of NSA. Section \uppercase\expandafter{\romannumeral6}-A introduces the HPC cluster network in which all instances are based along with the parameter settings. Section \uppercase\expandafter{\romannumeral6}-B - \uppercase\expandafter{\romannumeral6}-E compare different NN architectures of RL, SA and NSA, SCIP and NSA, and different scheduling methods, respectively. All algorithms and simulations are programmed in Python and executed on a on server with 2 Intel Xeon 6226 CPUs and a single Nvidia Tesla V100-PCIE GPU. Finally, all software and test instances developed in this project, with which all results presented in this section can be replicated, are available open-source at https://github.com/lzk23/hpcschedule.

\subsection{HPC cluster network and parameter setting}

The HPC cluster network structure is a 1000-node pruned fat-tree with a radix $k=20$. The workload consists of ${10}^5$ jobs, which are generated randomly with a requested number of nodes in [1, 40] and a processing time in [10, 1800]. If no special explanation, other parameters are set as Table \uppercase\expandafter{\romannumeral1} shows. 

\begin{table*}[t]
	\begin{center}
		\caption{The parameter settings.}
		\label{tab:1}
		\def\arraystretch{1.2}
		\begin{tabular}{c c c}
			\hline
			\textbf{Parameter} & \textbf{Definition} & \textbf{Value}\\
			\hline
			$T^{\mathrm{max}}\left(T^{\mathrm{min}}\right)$  &  Maximal (Minimal) temperature of SA and NSA    & 2500(2.5)       \\
			$t^{\mathrm{max}}$   &  Maximal iteration steps of SA and NSA       & 500        \\
			$c$ &  One CH cost    & 1000        \\ 
			$n^D$ &  Maximal remove number of jobs in remove operator     & 2        \\ 
			$q$ &  Constant parameter used in state representation    & 100        \\ 
			$\tau$ &  Length of each time period    & 60s        \\
			\hline
		\end{tabular}
	\end{center}
\end{table*}

\subsection{Comparison for the NN architectures of RL}
This section compares the RL algorithms based on two different types of NN architectures, namely FCN and CNN, detailed in Section \uppercase\expandafter{\romannumeral5}-B. Further, as shown in Table \uppercase\expandafter{\romannumeral2}, three NN architectures with different network sizes (expressed as the number of parameters) are designed based on FCN and CNN, respectively. The FCN-based NNs consists of fully-connected (Linear) layers \cite{2020Impact}, while CNN-based NNs consists of one-dimensional convolutional layers \cite{MD2014Visualizing} and fully-connected layers. In particular, there is a process of flattening (Flatten) that converts multidimensional data into one-dimensional data between convolutional layers and fully-connected layers in CNN-based NNs. Tabh \cite{Mirhassani2014Efficient} is used as activation function. Remember that PPO implements an actor-critic architecture. Table \uppercase\expandafter{\romannumeral2} merely gives the information relative to actor models. The NN architectures of actor model and critic model usually only differ in the last layer. For example, for the first FCN-based NN, the corresponding NN architecture of critic model can be derived by using “Linear(64,1)” instead of “Linear(64,100)” to output only one critic value.

Fig. 4 compares the learning curves for RL with the six NNs, where the curves are smoothed by moving averages to improve readability. For notational convenience, the three FCN-based NNs are represented as FCN-1, FCN-2, and FCN-3, respectively. Similarly, the three CNN-based NNs are represented as CNN-1, CNN-2, and CNN-3, respectively. It can be seen that the performances between CNN-1, CNN-2, and CNN-3 have a clear difference, whereas it is not between FCN-1, FCN-2, and FCN-3. In addition, the CNN-based models are better than the FCN-based model. Even the worst-performing CNN-based model (i.e., CNN-1) has a better performance than that of the best-performing FCN-based model (i.e., FCN-3). Although the number of parameters of CNN-1 is less than that of FCN-3.

\begin{table*}[t]
	\begin{center}
		\caption{The NN architectures of RL.}
		\label{tab:1}
		\def\arraystretch{1.2}
		\begin{tabular}{c|c c c}
			\hline
			\textbf{NN architecture} & \textbf{1} & \textbf{2} & \textbf{3} \\
			\hline
			FCN  & \makecell{Linear(400, 64) \\ Tanh() \\ Linear(64,64) \\ Tanh() \\ Linear(64, 100)}  
			& \makecell{Linear(400, 64) \\ Tanh() \\  Linear(64,64) \\ Tanh() \\ Linear(64,64) \\ Tanh() \\ Linear(64, 100)}  
			& \makecell{Linear(400, 96) \\ Tanh() \\ Linear(96,64) \\ Tanh() \\ Linear(64, 100)} \\ 
			\hline
			\textbf{\# of parameters} & 42724 & 46884 & 60804 \\
			\hline
			CNN & \makecell{Conv1d(4, 4, 10, 1,0) \\ Tanh() \\ Conv1d(4,4,5,2,0) \\ Flatten() \\ Linear(176,256) \\ Linear(256,100) }
			& \makecell{Conv1d(4, 8, 10, 1,0) \\ Tanh() \\ Conv1d(8,4,5,2,0) \\ Flatten() \\ Linear(176,256) \\ Linear(256,100) }
			& \makecell{Conv1d(4, 16, 10, 1,0) \\ Tanh() \\ Conv1d(16,8,5,2,0) \\ Flatten() \\ Linear(352,256) \\ Linear(256,100)}\\ 
			\hline
			\textbf{\# of parameters} & 37248 & 42097 & 54641 \\	        
			\hline
		\end{tabular}
	\end{center}
\end{table*}

\begin{figure}[t]
	\centerline{
		\includegraphics[width=1\columnwidth]{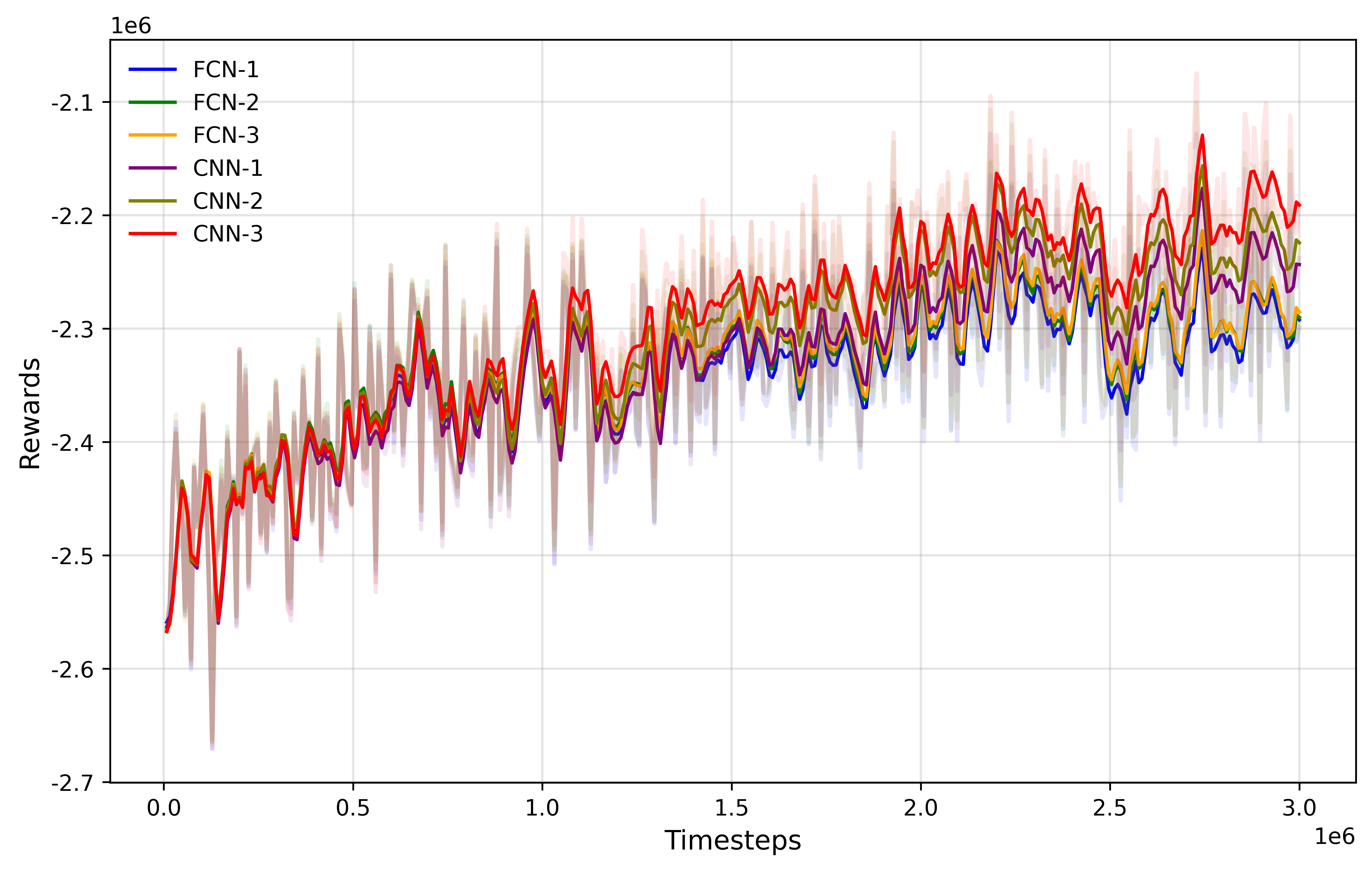}}
	\caption{Comparisons of the learning curves of RL with different model architectures.}
	\label{fig:repair_problem}
\end{figure}

\subsection{Comparison of SA and NSA}
In this section, NSA and SA will be compared in order to assess the improvement of NSA over SA. The NN architecture of RL used in the repair operator of NSA is CNN-3. The comparison consists of 10 epochs, where each epoch contains 100 dynamically generated instances. Specifically, all nodes in the HPC cluster are initialized as idle before the beginning of each epoch. The jobs appear continuously as time goes by. At each decision period, the TJAP problem is solved by NSA and SA, respectively. Meanwhile, the states of waiting queue and cluster are updated according to the result obtained from NSA, which ensures that the NSA and SA solve the same instance. For notational convenience, NSA(SA) with a special iteration number $t^{\mathrm{max}}$ is denoted as NSA-$t^{\mathrm{max}}$(SA-$t^{\mathrm{max}}$). For example, “SA-500” refers to the SA algorithm with 500 iterations. Table \uppercase\expandafter{\romannumeral3} shows the comparison results, where “Cost” and “Time” columns refer to the average CH cost and average solving time of 100 instances in each epoch, respectively; “Seq.” is the initial solution derived by a sequential assignment that schedules jobs one by one based on the SCAS; “Impr.” indicates the percentage improvement of solution quality achieved by NSA-500 in relation to SA-1000. Due to the embedding of NN calculation in NSA, the calculation time of NSA is longer than that of SA with the same iteration steps. To assess the performance improvement of NSA compared with SA when the computation times are same, Table \uppercase\expandafter{\romannumeral3} provide the results of SA-1000 and “Impr.”.

\begin{table*}[t]
	\centering
	\setlength{\abovecaptionskip}{2pt}%
	\setlength{\belowcaptionskip}{10pt}%
	\caption{Comparisons between the results of SA and NSA.}
	\begin{tabular}{c|ccccccccc}
		\bottomrule 
		\multicolumn{1}{c}{\multirow{2}[4]{*}{\textbf{Epoch}}} & \multicolumn{1}{c}{\textbf{Seq.}} & \multicolumn{2}{c}{\textbf{SA-500}} & \multicolumn{2}{c}{\textbf{SA-1000}} &  \multicolumn{2}{c}{\textbf{NSA-500}}  & \multicolumn{1}{c}{\multirow{2}[4]{*}{\textbf{Impr.}}} \\
	    \cmidrule(r){2-2}  \cmidrule(r){3-4} \cmidrule(r){5-6} \cmidrule(r){7-8}  \multicolumn{1}{c}{} & \multicolumn{1}{c}{Cost} & \multicolumn{1}{c}{Cost} & \multicolumn{1}{c}{Time/s} & \multicolumn{1}{c}{Cost} & \multicolumn{1}{c}{Time/s} & \multicolumn{1}{c}{Cost} & \multicolumn{1}{c}{Time/s} & \multicolumn{1}{c}{} \\
		\hline
		1 &	99152 &	87322 &0.8 &	86958 &1.5	 & 85596 &	1.5	 & 1.57\% \\
		2 &	105087 & 91662	&  0.7  &	90810 &	 1.5 &	89026 &	1.5	 &  1.96\% \\
		3 &	110728 &	97723&	0.8&	97006&	1.6	&96215	&1.6	&0.82\% \\
		4	&102931&	90909&	0.9&	89797&	1.9	&89127&	1.9&	0.75\% \\
		5	&105780&	94016&	0.7&	92483&	1.5&	91471&	1.5&	1.09\% \\
		6	&106006&	91388&	0.7&	90084&	1.5&	88919&	1.5&	1.29\% \\
		7	&105448&	93022&	0.7&	91804&	1.5&	91060&	1.5&	0.81\% \\
		8	&104471&	84483&	0.8&	83266&	1.6&	82668&	1.4&	0.72\% \\
		9	&107885&	97214&	0.8	&96337&	1.5&	94642&	1.5&	1.76\% \\
		10	&102747&	90393&	0.7&	89552&	1.4&	87163&	1.5&	2.67\% \\ \midrule
		\textbf{Avg.}&	105023.5&	91813.2&	0.76&	90809.7&	1.55&	89588.7&	1.54&	1.34\% \\
		\toprule 
	\end{tabular}%
	\label{tab:3}%
\end{table*}%

Table \uppercase\expandafter{\romannumeral3} shows that the solving times of NSA-500 and SA-1000 are roughly equivalent. Therefore, we choose to compare NSA-500 with SA-1000. It can be seen from the “Impr.” column that NSA-500 has a better performance than SA-1000 in all epochs. The average improvement for the ten epochs is 1.34\%.
To reflect the improvement of NSAs with a repair operator based on different NNs (i.e., the six NNs shown in Table \uppercase\expandafter{\romannumeral2}) compared with SA, we use NSAs with different NNs and SA to solve one thousand instances equally distributed into ten epochs, respectively. Further, the number of iterations of NSA is set to 500, 600, 700, 800, 900, and 1000, respectively. To keep the solving time consistent, the number of iterations of SA is set to 1000, 1200, 1400, 1600, 1800, and 2000, respectively. Fig. 5 compares the average CH cost across six NNs and six different number of iterations. Although NSA-500 with FCN-1 has a worse performance than SA-1000 (the corresponding improvement is -0.27\%), in all other instances NSAs have lower CH costs (the improvements are positive). Especially, using CNN-3 network has a larger improvement than using other NNs.

\begin{figure*}[t]
	\centerline{
		\includegraphics[width=1.5\columnwidth]{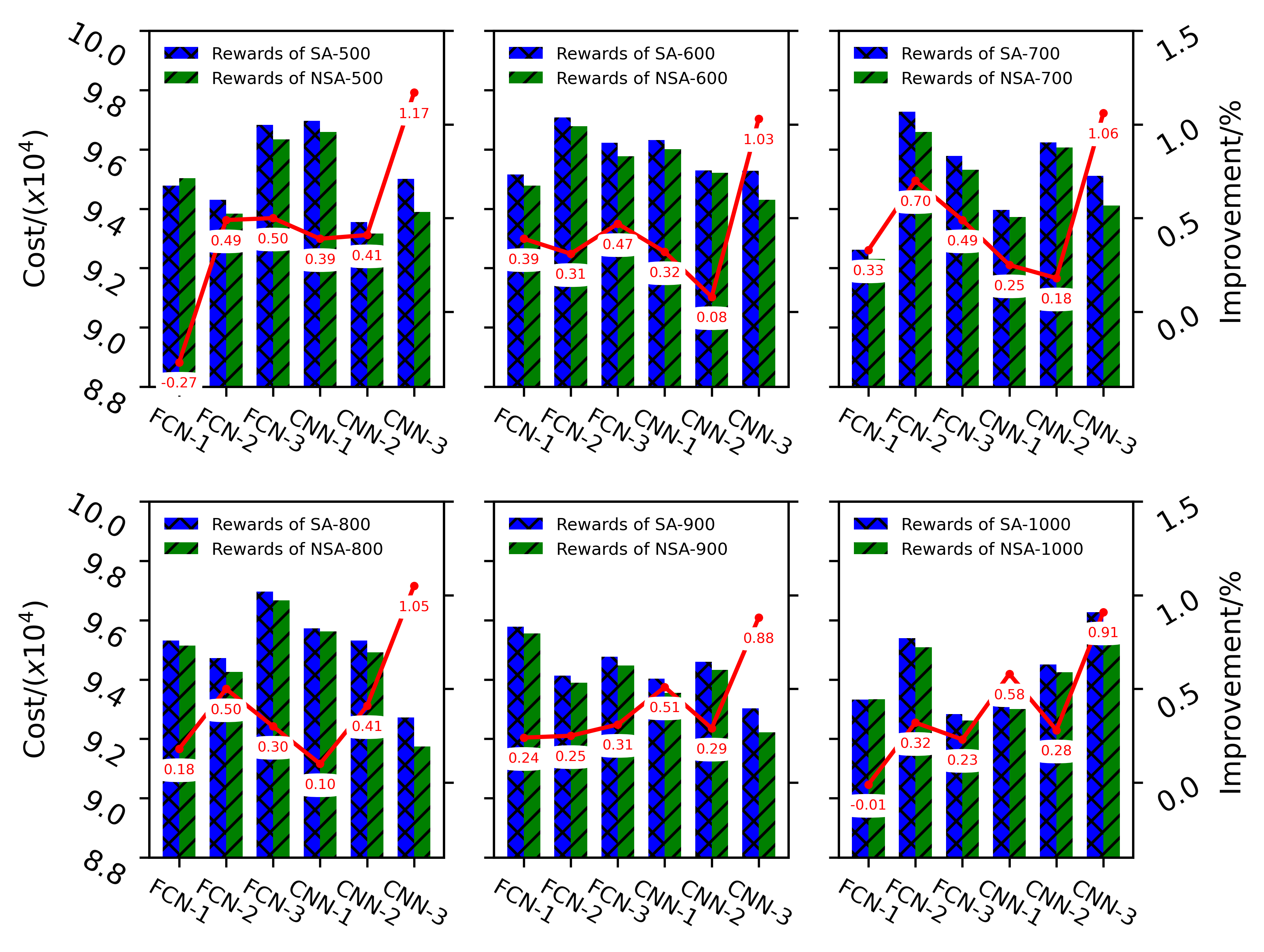}}
	\caption{Comparisons between NSAs with different NNs and SA.}
	\label{fig:repair_problem}
\end{figure*}

\subsection{Comparison of SCIP and NSA}

To reflect the advantages of NSA algorithm over the solver SCIP, the instances dynamically generated for the comparison between SA and NSA (see Table \uppercase\expandafter{\romannumeral3}) are also taken as input into model $\mathrm{M}_{\mathrm{SCAS}}$ and solved by SCIP. The results are shown in Table \uppercase\expandafter{\romannumeral4}, where “SCIP (no time limit)” means that the solving time of SCIP is unlimited; “SCIP (limit time=30s)” means that the solving time of SCIP is limited to 30s; “Cancel” refers to the percentage of unassigned jobs in each epoch to the selected jobs. Sometimes, although there is sufficient resource allocation, some jobs are not allocated because SCIP does not get a feasible solution in a limited time. “Impr.” indicates the percentage improvement of solution quality achieved by NSA in relation to “SCIP(no time limit)”. 

\begin{table*}[ht]
	\centering
	\setlength{\abovecaptionskip}{2pt}%
	\setlength{\belowcaptionskip}{10pt}%
	\caption{Comparisons the results of NSA and SCIP.}
	\begin{tabular}{c|ccccccccc}
		\bottomrule 
		\multicolumn{1}{c}{\multirow{2}[4]{*}{\textbf{Epoch}}} & \multicolumn{2}{c}{\textbf{SCIP(no time limit)}} & \multicolumn{3}{c}{\textbf{SCIP(time limit=30s)}} &  \multicolumn{2}{c}{\textbf{NSA}}  & \multicolumn{1}{c}{\multirow{2}[4]{*}{\textbf{Impr.}}} \\
		\cmidrule(r){2-3} \cmidrule(r){4-6} \cmidrule(r){7-8}  \multicolumn{1}{c}{} & \multicolumn{1}{c}{Cost} & \multicolumn{1}{c}{Time/s} & \multicolumn{1}{c}{Cost} & \multicolumn{1}{c}{Cancel} & \multicolumn{1}{c}{Time/s} & \multicolumn{1}{c}{Cost} & \multicolumn{1}{c}{Time/s} & \multicolumn{1}{c}{} \\
		\hline
		1&	79597&	26.2&	79484&	0.01&	8.5&	85596&1.6&	-7.54\% \\ 
		2&	83863&	14.7&	81421&	0.04&	8.5&	89026&	1.5&	-6.16\% \\
		3&	89422&	21.2&	86687&	0.04&	9.2&	96215&	1.6&	-7.60\% \\
		4&	82598&	21.5&	82410&	0.02&	8.7&	89127&	1.9&	-7.90\%\\
		5&	84438&	22.0&	80421&	0.03&	8.1&	91471&	1.5&	-8.33\%\\
		6&	80905&	16.6&	80380&	0.01&	7.9&	88919&	1.5&	-9.91\%\\
		7&	84358&	19.5&	85253&	0.01&	8.1&	91060&	1.5&	-7.94\%\\
		8&	76331&	12.2&	75734&	0.01&	6.3&	82668&	1.4&	-8.30\%\\
		9&	88091&	22.6&	87345&	0.02&	8.5&	94642&	1.6&	-7.44\%\\
		10&	81320&	13.1&	82013& 0.01&	7.9&	87163&	1.5&	-7.19\%\\  \midrule
		\textbf{Avg.}&	83092.3&	18.96&	82114.8&	0.02&	8.17&	89588.7&	1.56&	-7.82\%\\
		\toprule 
	\end{tabular}%
	\label{tab:3}%
\end{table*}%

Table \uppercase\expandafter{\romannumeral4} shows that the efficiency of SCIP cannot meet the actual requirements. When the solving time is limited to 30 seconds, some instances cannot obtain a feasible solution, that is, there are some jobs that have not been allocated. When the solving time of SCIP is unlimited, the average solving time is up to 18.96s, which is a far too long response time when considering real-time application. According to our observation, the solving time of SCIP is closely related to the number of idle nodes and the number of jobs allocated. On the contrary, in all epochs, the average solving time of NSA is less than 2s. Concerning solution quality (see column “Impr.”), the quality obtained by NSA is inferior to that of “SCIP(no time limit)”. Nonetheless, the quality difference is not more than 10\% in all epochs.

\subsection{Comparison of different scheduling methods}
This section compares the window-based scheduling method with the per-job manners, i.e., FCFS and BF. In addition, the length of each time period in the window-based scheduling method is set to 10s, 20s, 30s, 40s, 50s and 60s respectively to analyze the impact on CH costs and waiting time (the allocation time minus the arrival time). Each method is tested over 10 epochs. In each epoch, 300 continually arriving jobs are allocated by using different methods respectively. Fig. 6 shows the results. It should be pointed out that the CH cost will affect the processing time of a job, thus affecting the average waiting time. However, the specific relationship between CH cost and running time is not within the scope of this study, so it is still assumed that the processing times of job does not change with the change of CH cost.

\begin{figure}[t]
	\centerline{
		\includegraphics[width=1\columnwidth]{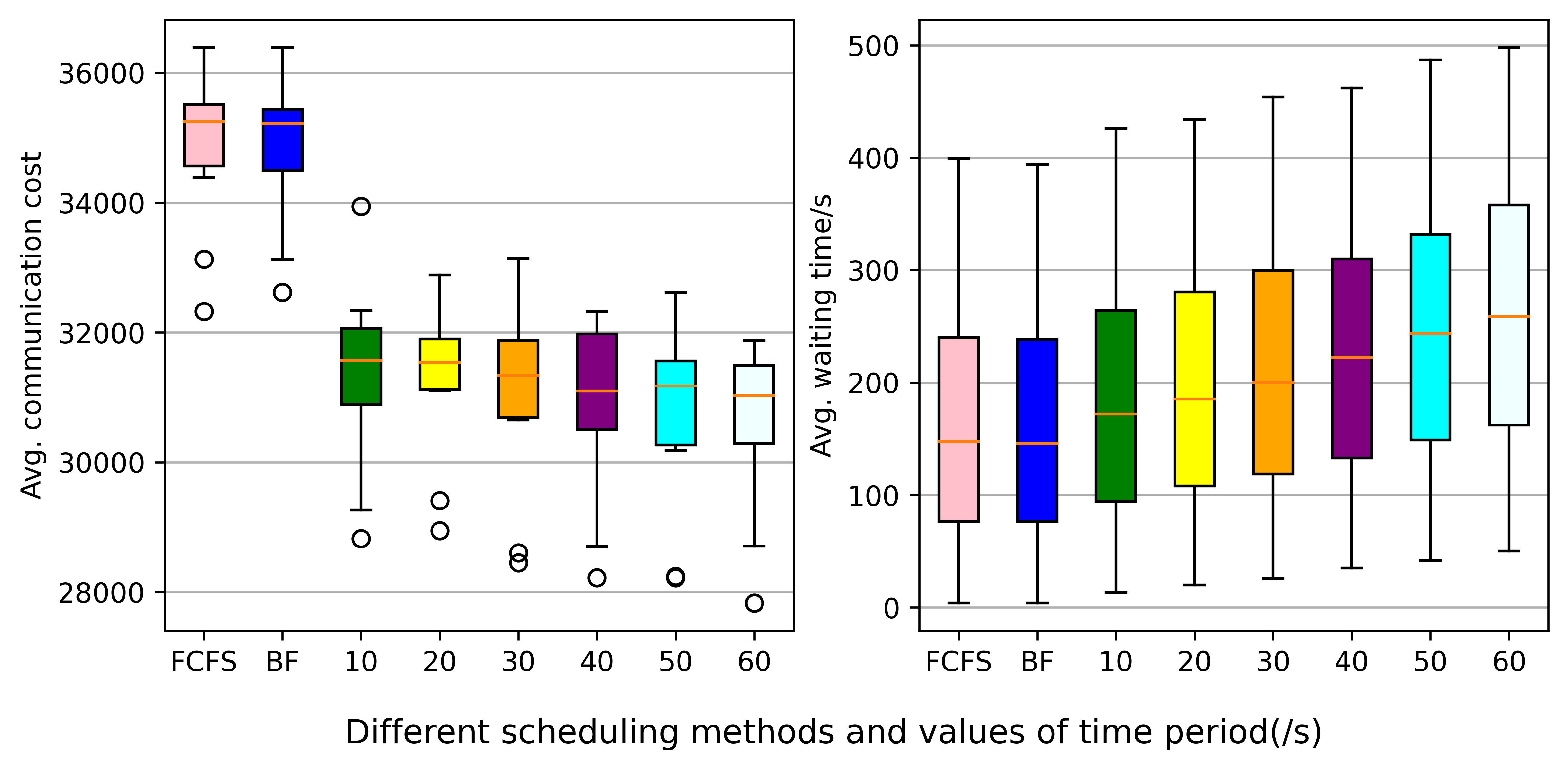}}
	\caption{Comparisons the results of different scheduling methods and lengths of time period.}
	\label{fig:repair_problem}
\end{figure}

Fig.6 indicates that compared with FCFS and BF methods, the window-based scheduling method can reduce the average CH cost at the cost of the average waiting time. In addition, as the length of each time period increases, the average CH cost tends to decrease, while the average waiting time increases.

\section{CONCLUSIONS}
In this paper, we study the window-based TJAP in a fat-tree based HPC system aiming at reducing the CH cost. The special SCAS and the DCAS are proposed to avoid the pitfall of dealing with QAP. For the SCAS, we develop a 0-1 integer programming which can be directly solved by a solver; whereas for the DCAS, a new algorithm, i.e., NSA, is proposed. Experiments show that, compared with SA, NSA can get better results with the same solving times. When using SCIP to solve model $\mathrm{M}_{\mathrm{SCAS}}$, we found that the solving time of SCIP is more sensitive to the solving scale and is unacceptable for the TJAP. Additionally, the solution quality of SCIP with a limited time 30s is obviously inferior to that of NSA. Compared with the per-job scheduling methods, namely FCFS and BF, the average CH cost can be reduced with some sacrifice of increasing the waiting time. 

\section*{Acknowledgements}
This work was supported in part by China Postdoctoral Science Foundation grant under the contract 2022M720931.

\bibliographystyle{IEEEtran}
\bibliography{ref}

\end{document}